\documentclass[]{aastex631}
\usepackage{booktabs}
\usepackage{longtable}
\usepackage{amssymb}
\usepackage{placeins}

\begin{document}


\title{Gaia in Technicolor. II. Overlooked Post-Common-Envelope Hot Subdwarf \\ Binary Candidates Identified from Their Variability-Encoded CMD Trajectories}

\correspondingauthor{Bennett C. Kirby}
\email{kirbyben@unc.edu}

\author[0009-0003-4021-0934]{Bennett C. Kirby}
\affiliation{Department of Physics \& Astronomy, University of North Carolina at Chapel Hill \\
120 E. Cameron Ave. Phillips Hall\\
Chapel Hill, NC 27599, USA}

\author[0000-0002-8558-4353]{Brad N. Barlow}
\affiliation{Department of Physics \& Astronomy, University of North Carolina at Chapel Hill \\
120 E. Cameron Ave. Phillips Hall\\
Chapel Hill, NC 27599, USA}

\author[0009-0003-8885-523X]{Violet Trbonja}
\affiliation{Department of Physics \& Astronomy, University of North Carolina at Chapel Hill \\
120 E. Cameron Ave. Phillips Hall\\
Chapel Hill, NC 27599, USA}

\begin{abstract}
Hot subdwarfs are low-mass, core helium-burning objects that likely formed through binary interactions when their main sequence progenitors ascended the red giant branch. They serve as important probes of binary evolution and can be progenitors of Type Ia supernovae and gravitational wave emitters. Most hot subdwarfs in compact binaries show light curve variations that can be modeled to determine fundamental properties of both the stars and the system as a whole. This makes photometric surveys like Gaia, TESS, and ZTF invaluable to studies of these objects. However, searches for new hot subdwarf binaries typically begin with catalogs of hot subdwarf candidates that, by design, are filtered by astrometric, photometric, and crowding cuts, leaving out many potentially interesting systems. Gaia DR3 contains multi-epoch photometry of 11.7 million sources, each with about 50 measurements in three filters taken over 1,000 days. In Paper I of this series, we used these near-simultaneous $\mathrm{G_{BP}}$, G, and $\mathrm{G_{RP}}$ measurements to construct a variability-encoded color-magnitude diagram (VECMD), in which each type of variable traces a distinct trajectory through the CMD. Because the shape of this trajectory does not depend on the precision of a star's parallax, it can be used to find and classify variable hot subdwarfs that catalogs might leave out. Here we search 5,782 sources with multi-epoch photometry that lie in the hot subdwarf region of the Gaia CMD but were excluded from the widely used catalogs of \citet{Culpan2022}. Rather than relying on the astrometric data used to generate candidate catalogs, we preliminarily classify objects solely based on the shapes of their CMD variability trajectories and phase-folded light curves. With these data, we identified 27 new compact hot subdwarf binary candidates, 15 of which are eclipsing HW Vir-type systems, and 12 of which are new candidate reflection effect systems. Since 23 of these had never been identified as hot subdwarfs, they are newly classified candidate hot subdwarf binaries. Spectroscopic follow-up is needed to confirm their nature. Studies of these systems may further improve our understanding of the common envelope processes that formed them. We are working to extend our methodology to search for hot subdwarfs in other regions of the VECMD and to build machine-learning models to discover and classify systems efficiently based only on their CMD trajectory shapes.
\end{abstract}



\section{Introduction} \label{sec:intro}

Hot subdwarf B (sdB) stars were formerly red giant stars that were stripped of their outer H atmospheres near the tip of the red giant branch due to gravitational interactions with a nearby companion. With temperatures and log g values ranging from 20,000 -- 40,000 K and 5.0 -- 6.0 dex, respectively, they lie on the extended horizontal branch (EHB) of the Hertzsprung--Russell (H--R) diagram and are consistent with core He-burning objects with lifetimes around 100 Myr. A variety of binary formation channels have been proposed, invoking Roche lobe overflow and common envelope evolution processes that can produce sdB binaries with periods ranging from less than 1 hour to several years \citep{Han2002, Han2003}. Observations generally support the binary formation hypothesis, with some claiming that {\em all} hot subdwarfs are formed through binary interactions \citep{Pelisoli2020}. Approximately one-third of sdBs reside in close binaries with white dwarf (WD), brown dwarf (BD), or red dwarf (dM) companions, and orbital periods ranging from hours to days \citep{Maxted2001, Kupfer2015, Schaffenroth2019, Schaffenroth2022}. These compact systems show a rich variety of photometric variability, including pulsations, reflection effects (irradiation), eclipses, ellipsoidal deformation, and relativistic beaming. Together, these effects produce distinctive light curve morphologies that can be used to classify each binary subtype. Because most compact sdB binaries are photometrically variable, their light curves are often the first and most efficient way to discover them, and modeling these variations reveals the properties of both stars and the evolutionary channels that formed them.

Compact sdB binaries offer insights into diverse subfields of astrophysics. Some sdB+WD binaries are candidate Type Ia supernova progenitors \citep{WangHan2010, Mattig2026}, with combined stellar masses in excess of the Chandrasekhar limit \citep{Pelisoli2021}. Other similar binaries might also produce Type Ia supernovae that are below this limit, via the proposed double-detonation scenario \citep{Vennes2012, Geier2013}. Due to their short orbital periods and moderate masses, several sdB+WD systems are also classified as detectable gravitational wave sources for ESA's upcoming LISA mission \citep{Kupfer2024}. The frequency and amplitude of these gravitational waves should be at or above LISA's detection limit after 48 months of observation, and so these systems serve as important verification sources for the mission. Additionally, many compact sdB+dM/BD binaries are eclipsing and show a strong reflection effect in their light curves. Using time-series spectroscopy and photometry, modeling these binaries (e.g., \citealt{Schaffenroth2015}) can help constrain the common-envelope evolution channels that formed them. The resulting mass and period distributions help demonstrate the range of companion masses that can strip a red giant to form an sdB. Finally, many sdBs exhibit pulsations \citep{Kilkenny1997, Green2003} that reveal a wealth of information about their interiors \citep{LynasGray2021}.

\begin{figure*}[h!]
\centering
\begin{minipage}[t]{0.32\textwidth}
    \centering
    \includegraphics[width=\linewidth]{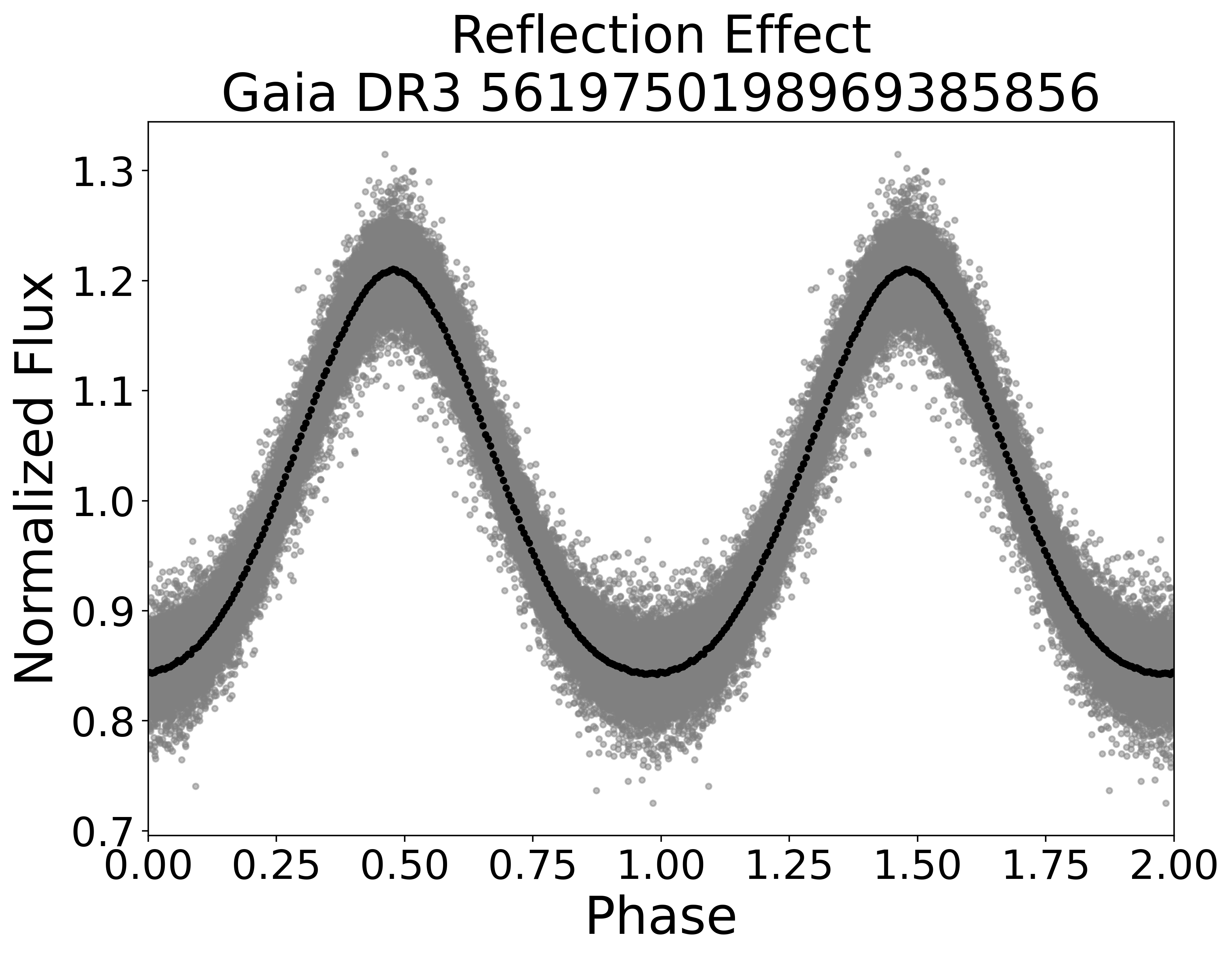}

    \smallskip
    \textbf{\ \ \ (a)}
\end{minipage}\hfill
\begin{minipage}[t]{0.32\textwidth}
    \centering
    \includegraphics[width=\linewidth]{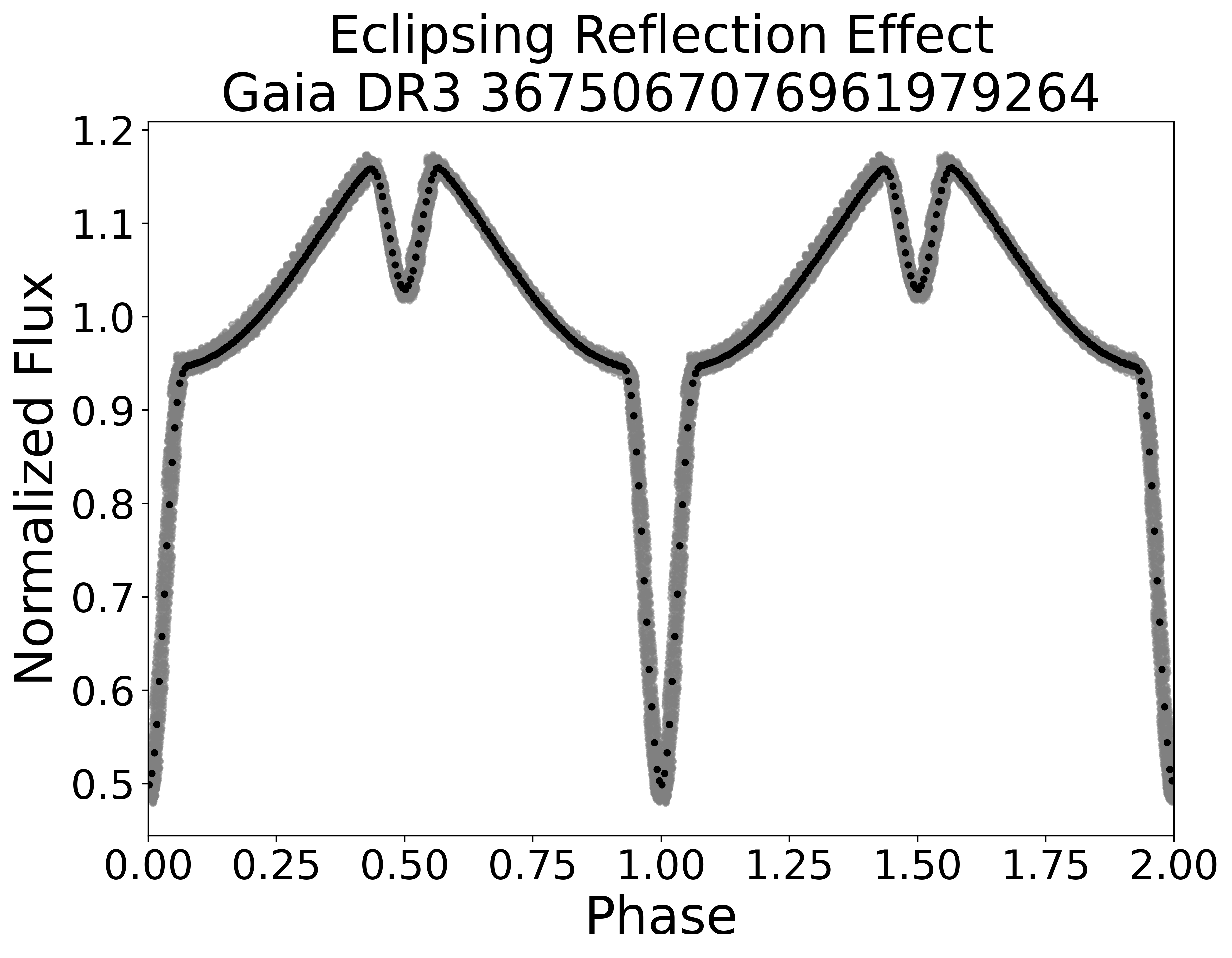}

    \smallskip
    \textbf{\ \ \ (b)}
\end{minipage}\hfill
\begin{minipage}[t]{0.32\textwidth}
    \centering
    \includegraphics[width=\linewidth]{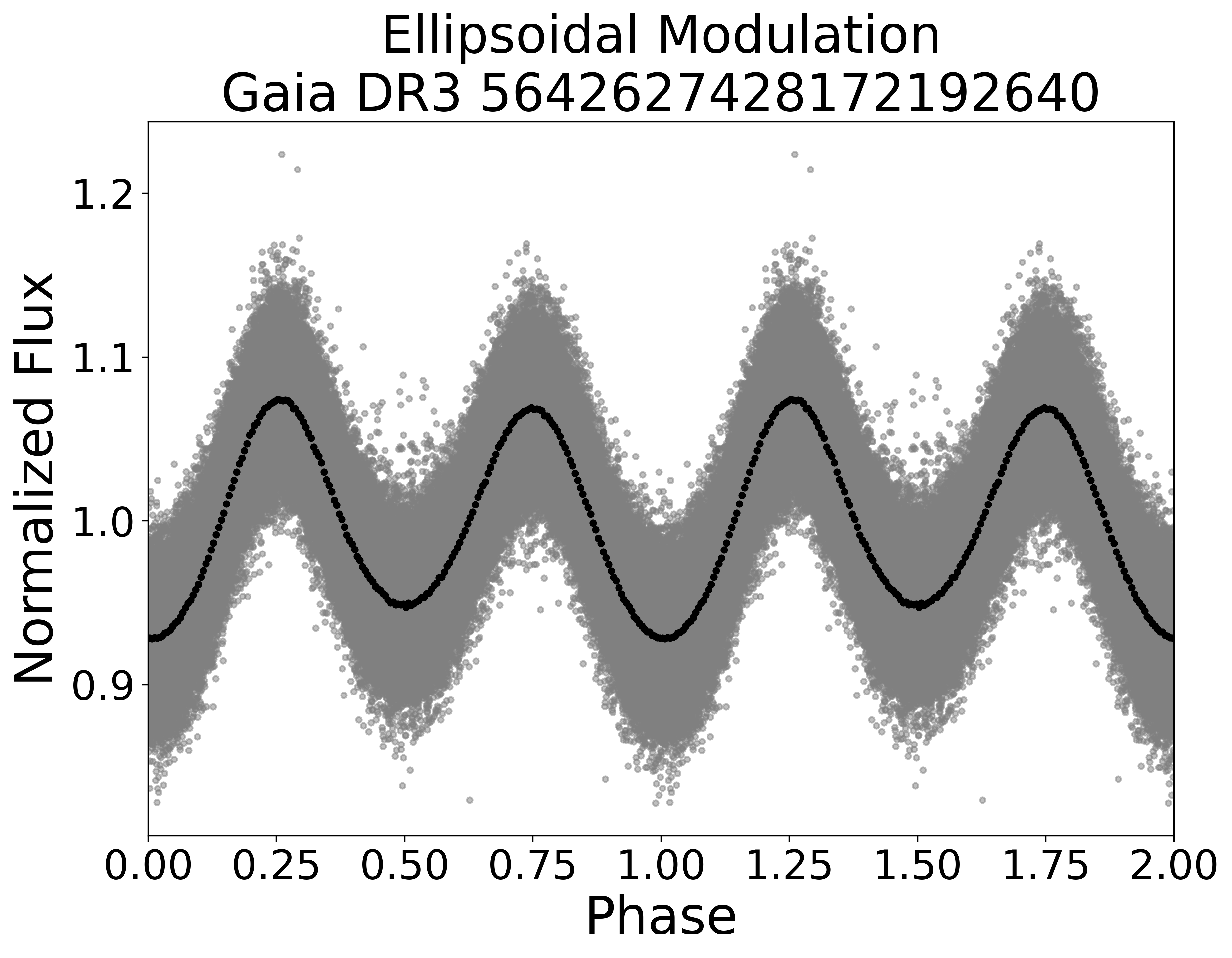}

    \smallskip
    \textbf{\ \ \ (c)}
\end{minipage}

\caption{Examples of the three main light curve morphologies observed in compact sdB binaries. The data shown are from TESS and phase-folded on published periods given by \citet{Krzesinksi2022}, \citet{Kiss2000}, and \citet{Ratzloff2020} for systems (a), (b), and (c), respectively. The TESS Input Catalog (TIC) IDs associated with systems (a), (b), and (c) are TIC 5051080, TIC 156618553, and TIC 1973623. The gray points show all of the short-cadence data released by TESS, and the  black points show binned data. \textbf{(a)} The reflection effect sdB+dM system TIC 5051080 has a flatter trough and a sharper peak caused by the irradiated dayside of the dM coming in and out of view. \textbf{(b)} The HW Vir-type system TIC 156618553 shows eclipses in addition to its own reflection effect, with a $\sim$$50\%$ dip in brightness during the primary eclipse and a $\sim$15\% dip during the secondary eclipse. Overall, the quasi-sinusoidal reflection effect shape remains evident in the light curve when the system is not eclipsing. \textbf{(c)} The ellipsoidal sdB+WD system TIC 1973623 with a quasi-sinusoidal shape shows two cycles of brightening/dimming for one orbit due to the geometry of the system. Every other trough is deeper due to gravity darkening, and every other peak is higher due to relativistic beaming.}
\label{fig:tess_lcs}
\end{figure*}

Many past surveys and instruments have collected high-cadence optical photometry of sdBs. One of the most influential is the Transiting Exoplanet Survey Satellite (TESS), which has aided in the classification of hundreds of variable sdBs using its short-cadence data mode (e.g., \citealt{Barlow2024, Schaffenroth2023}). As presented in Figure \ref{fig:tess_lcs}, three main light curve morphologies tend to dominate such surveys when looking at compact hot subdwarf binaries. Figure \ref{fig:tess_lcs}(a) shows a reflection effect light curve indicative of short-period sdB+dM/BD systems. In these binaries, the dayside of the cool companion facing the sdB is heated to 10,000--20,000 K, and the irradiation produces strong, quasi-sinusoidal photometric variations as the heated face comes in and out of view. The flatter trough of the quasi-sinusoid and sharper crest are characteristic of reflection effect systems, which typically have peak-to-peak amplitudes around 10--20\%. Figure \ref{fig:tess_lcs}(b) shows an eclipsing reflection effect sdB+dM/BD binary. These ``HW Vir'' systems display the quasi-sinusoidal modulations as seen in compact reflection effect binaries, while also exhibiting primary eclipses at the troughs and secondary eclipses at the crests. The primary and secondary eclipses can vary greatly in depth and duration depending on the inclination of the system, the size of the companion, and the orbital period. Figure \ref{fig:tess_lcs}(c) shows the light curve morphology for a compact sdB+WD binary displaying ellipsoidal modulation due to tidal deformation of the sdB. These quasi-sinusoidal light curves show two cycles of brightening/dimming per orbital period caused by the teardrop shape of the ellipsoidal sdB. The two troughs are often asymmetric in their depths because of gravity darkening, while the two crests can peak at different heights as a consequence of relativistic beaming. The peak-to-peak amplitude from ellipsoidal modulation can be as high as $\sim$15\%. Although TESS has been instrumental in discovering, characterizing, and monitoring compact sdB binaries, its small aperture size generally limits studies to bright objects ($T<16.5$ mag). 

The European Space Agency launched Gaia in 2013 with the goal of creating the most precise three-dimensional map of the Milky Way \citep{Gaia2016b}. To accomplish this goal, Gaia collected astrometry and photometry for approximately 2 billion sources over a little more than a decade \citep{Gaia2023j}. Color-magnitude diagrams (CMDs) created from averaged $\mathrm{G_{BP}}$, $\mathrm{G}$, and $\mathrm{G_{RP}}$ magnitudes from Gaia Data Release 2 (DR2) led to the discovery of tens of thousands of new candidate sdBs, increasing the number of known sdBs by a factor of about 10 \citep{Geier2019}. This new collection of candidate sdBs has provided a rich data set for follow-up studies that continue to improve our understanding of late stages of stellar evolution. Past studies have successfully used Gaia DR2 time-averaged photometry to identify new variable stars through anomalously large photometric uncertainties, without accessing the individual epochs of targets (e.g., \citealt{Barlow2022,Guidry2021}).

The largest catalog of sdB stars was published by \citet{Culpan2022} after their analysis of time-averaged Gaia EDR3 photometry. \citet{Culpan2022} provided a list of 61,585 candidate sdBs in Gaia EDR3 and also released a catalog of known hot subdwarfs containing 6,616 systems of various subtypes. These catalogs have become the standard starting point for discovering and characterizing new hot subdwarf binaries (e.g., \citealt{Princy2025}). However, to ensure a clean sample, \citet{Culpan2022} selected candidates with parallax uncertainties below 20\% in a CMD cut and the rest by reduced proper motion, applying astrometric, photometric and crowding cuts. These types of cuts are needed when a star's classification relies on its position in the CMD, but they inevitably leave out many bona fide hot subdwarfs, including variable hot subdwarfs. Since these systems are absent from standard catalogs, they are also largely absent from searches for new variable stars, and many interesting binaries likely remain undiscovered.

The Gaia Collaboration released multi-epoch photometry (MEP) across all three filters for over 11.7 million sources in DR3 \citep{GaiaVar}. The photometry covers a baseline of approximately three years, with most sources having about 50 brightness measurements per filter taken somewhat randomly over this time period. Importantly, the three filter observations at each observing epoch are essentially contemporaneous, with the total transit window being roughly 40--50 seconds. Despite the sparse sampling of the data, variability can still be detected in many objects due to Gaia's impressive photometric precision. Past studies have used machine learning algorithms along with phase-folded light curves to identify and classify variable sdBs using Gaia DR3 MEP (e.g., \citealt{Princy2025}).

In Paper I of this series (\citealt{PaperI}; hereafter, Paper I), we introduced a simple variability-encoded color-magnitude diagram (VECMD) in which each star is represented not by a single, static point but by a line segment representing the approximate path it traces in the CMD as it varies. Because the exact movement through the CMD is determined by the physics underlying its variability, each type of variable traces a unique CMD trajectory shape. Reflection effect binaries, for example, trace tight, tilted paths and are reddest when brightest. HW Vir-type systems additionally show eclipse features that drop down from the ends of this path. We note that the shape of a CMD trajectory does not depend on the precision of a star's parallax, nor does it change significantly with reddening. Thus, CMD trajectory shapes can be used to find and classify new variables, even when distances are poorly constrained.

In this paper, we use CMD trajectory shapes to search for new compact hot subdwarf binaries among the systems left out of the \citet{Culpan2022} catalog. In Section \ref{sec:MEP}, we present Gaia DR3 MEP data and CMD trajectories for known variable hot subdwarfs to demonstrate the utility of this approach. In Section \ref{sec:target}, we describe how we selected 5,782 systems that lie in the hot subdwarf region of the CMD but were excluded from the known and candidate catalogs. In Section \ref{sec:results}, we present 27 new candidate variable hot subdwarf binaries identified from these methods, most of which had never before been recognized as hot subdwarfs. Lastly, in Section \ref{sec:conclusion}, we discuss the spectroscopic follow-up needed to confirm these systems and future projects that will utilize machine learning to take advantage of all the new photometry coming in Gaia DR4.

\section{Gaia DR3 Multi-Epoch Photometry}\label{sec:MEP}

Gaia DR3 MEP tracks the brightness of objects with extremely high precision over about three years. The telescope took data in three filters, providing near-simultaneous measurements in  $\mathrm{G_{BP}}$, G, and $\mathrm{G_{RP}}$ \citep{Gaia2023j}. These data combined with parallaxes yield absolute G magnitude ($\mathrm{M_G}$) measurements, and a $\mathrm{G_{BP}-G_{RP}}$ color value for each epoch of data. Plotting each epoch on a CMD allows us to investigate the path the system takes through the CMD as it varies over time. Each sdB variable subtype changes brightness due to different physical processes described in Section \ref{sec:intro}. These physical processes produce distinct light curve morphologies, as shown in Figure \ref{fig:tess_lcs}. Therefore, different types of sdB variables should show distinct trajectories through the CMD, which can be represented in the VECMD and used to discover and classify new objects (Paper I). In this section, we demonstrate this approach using the three known sdB binaries from Figure \ref{fig:tess_lcs}.

\subsection{Light Curves and Periodograms}\label{subsec:LCs}

\begin{figure}[h!]
    \centering
    \includegraphics[width=\linewidth]{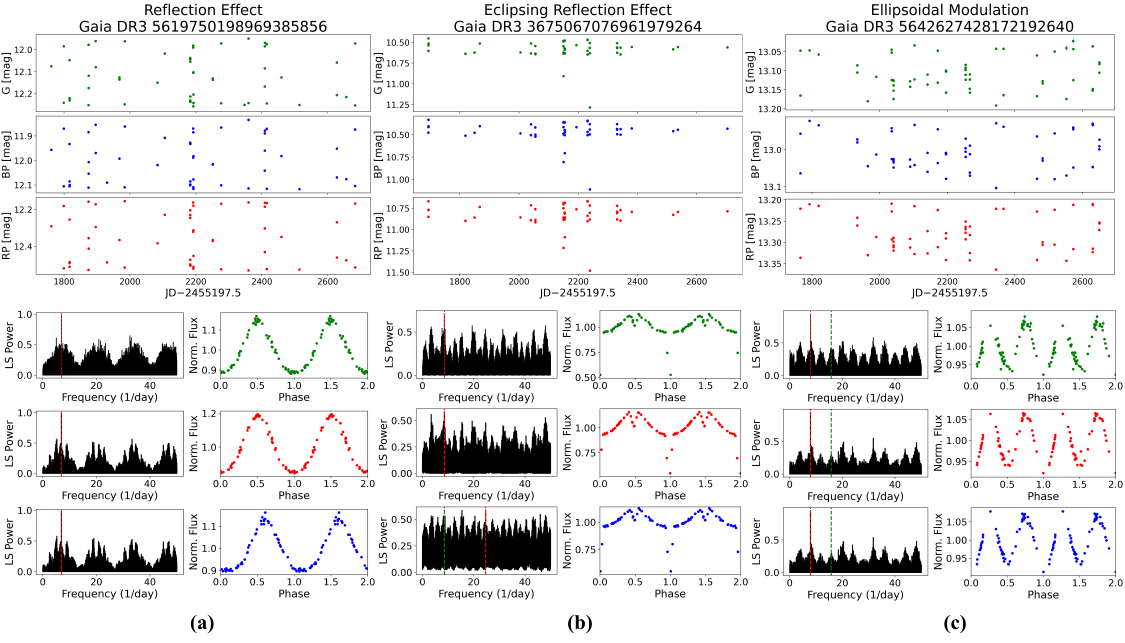}
    \caption{\textit{Top:} Examples of the full Gaia DR3 MEP data set for a reflection effect, an HW Vir-type, and an ellipsoidally modulated sdB. Systems \textbf{(a)}, \textbf{(b)}, and \textbf{(c)} are the same as in Figure \ref{fig:tess_lcs}. Green points represent the G filter, while the blue and red points are data from the $\mathrm{G_{BP}}$ and $\mathrm{G_{RP}}$ filters, respectively. About 50 epochs of data are available for each system and filter, with data sets spanning about 1,000 days. \textit{Bottom:} Periodograms run for each filter are beside the corresponding data phase-folded on the highest peak in the periodogram (or 2 times the highest peak for the ellipsoidal sdB light curve in panel \textbf{(c)} shown with the dashed green line). We match the shapes found in Figure \ref{fig:tess_lcs} for each system type. However, the strongest peak (dashed red line) in the periodogram for the HW Vir-type system's $\mathrm{G_{BP}}$ filter was not the known period provided by \citet{Kiss2000} (dashed green line). Here, we fold over the known period to demonstrate that the standard HW Vir-type light curve shape is present in the data. Note that each point shown in the top and bottom panels has error bars.}
    \label{fig:Gaia_mep}
\end{figure}

With three filters per observation epoch, Gaia DR3 MEP  allows us not only to look for variability within a single filter, but also {\em color} variability over time. In Figure \ref{fig:Gaia_mep}, we see the full Gaia DR3 MEP data for the same three systems shown in Figure \ref{fig:tess_lcs} with sigma-clipped MEP data in the top panel and the phase-folded light curves with corresponding periodograms in the bottom panel.

Compared to data from surveys like TESS and ZTF, Gaia DR3 MEP is a sparsely sampled data set, with each system being observed only $\sim$50 times over a three-year baseline. Nonetheless, Gaia's incredible photometric precision, combined with its relatively random sampling cadence, permits the detection of coherent signals with surprising success, even down to periods as short as minutes (Paper I). To phase-fold the Gaia DR3 MEP data in all stages of this project, we computed a Lomb-Scargle periodogram (LSP) for each filter using the \texttt{LombScargle} class from the \texttt{Astropy} package. Each LSP spanned from $10^{-2}$ to $300\ \mathrm{days}^{-1}$ (or a period of 4.8 minutes), safely encompassing the orbital periods of compact sdB binaries, and the periods of most sdBV pulsation modes. A periodogram's frequency resolution is given by $\Delta f = 1/\Delta T$, where $\Delta T$ is the time interval between the first and last observations. Therefore, the effective frequency resolution is $\Delta f \approx 10^{-3}\ \mathrm{days}^{-1}$ for each target. To ensure smooth sampling across the periodogram, we chose to oversample the frequencies by a factor of 10, which resolves all peaks while keeping computation quick. Occasionally, in one of the filters, known sdB binaries did not have a distinct peak in the calculated periodogram. We found that, in our experience, about $15\%$ of known sdB systems have at least one of the three filters with a periodogram whose largest peak was not consistent with the published period, likely due to aliasing. The G filter most often showed a distinct periodic signal because it has the largest bandwidth of the three filters, yielding the most precise photometry \citep{Gaia2023j}. Additionally, there were occasionally more points in the G filter than the $\mathrm{G_{BP}}$ or $\mathrm{G_{RP}}$ filters.

Despite the sparse data and cadence, the signals are easily picked up by the LSP for most systems, and the resulting phase-folded diagrams clearly and accurately represent the light curve morphologies of these known binaries. The reflection effect system in Figure \ref{fig:Gaia_mep}(a) shows the same quasi-sinusoidal shape as the TESS data in Figure \ref{fig:tess_lcs}(a). Similar agreement is found for the HW Vir system shown in Figure \ref{fig:Gaia_mep}(b), with Gaia having also caught the eclipses at multiple epochs. Finally, Figure \ref{fig:Gaia_mep}(c) shows that the light curve morphology of the ellipsoidally modulated sdB+WD binary from Figure \ref{fig:tess_lcs}(c) was also recovered. However, due to sparse sampling, the asymmetries in the troughs and crests are more difficult to detect.

Although the three filter measurements were not obtained at exactly the same time during each visit, we consider a group of $\mathrm{G_{BP}}$, G, and $\mathrm{G_{RP}}$ points as one ``epoch'' of contemporaneous measurements if the time of a given G measurement is within 0.005 days of the average time of the neighboring $\mathrm{G_{BP}}$ and $\mathrm{G_{RP}}$ measurements. This timescale is shorter than the orbital periods of all known compact hot subdwarf binaries. If this criterion is met, the $\mathrm{G_{BP}}$ and $\mathrm{G_{RP}}$ points are combined into a single $\mathrm{G_{BP}-G_{RP}}$ color and assigned to the corresponding G magnitude for each epoch. We adopt as the time for each epoch the average time from all three filter measurements. The average separation between the G and $\mathrm{G_{BP}-G_{RP}}$ measurements for our systems is 0.00036 days ($\sim$31 seconds), far shorter than the period of the fastest known HW Vir binary, which has a period of 61.2 minutes (C. Derbyshire et al. 2027, in prep.). The time between consecutive epochs for most of our systems ranged from 20--50 days, with some systems deviating from this range significantly in both directions. With the G and corresponding $\mathrm{G_{BP}-G_{RP}}$ measurements assigned for each epoch, we now explore the phase-dependent color information present in Gaia DR3 MEP for our systems.
 
\subsection{Color Curves}

Gaia DR3 MEP enables the classification of systems with unique light curve morphologies via the phase folding of a sparsely sampled data set, as shown in Section \ref{subsec:LCs}. Investigating the phase-folded data in Figure \ref{fig:Gaia_mep} further, one can see that the variation due to the reflection effect is stronger in the $\mathrm{G_{RP}}$ filter than it is in the $\mathrm{G_{BP}}$ filter for both the reflection effect and HW Vir-type systems. The ellipsoidal system, however, has approximately equal variations in all three filters. This suggests that Gaia can detect differences in the color variations of different types of variables that might help reveal their nature. 

\begin{figure}[h!]
    \centering
    \includegraphics[width=\linewidth]{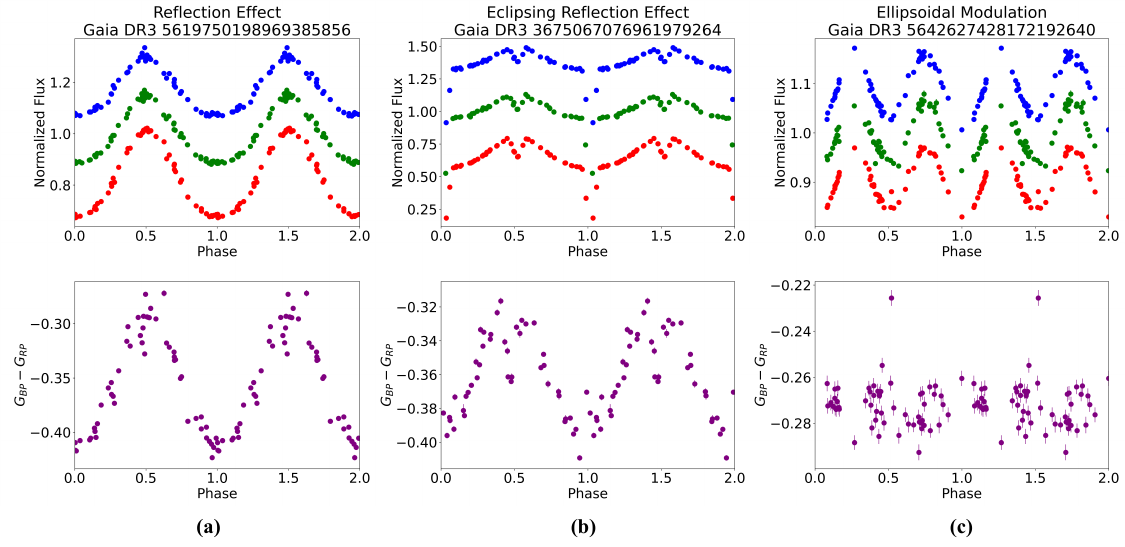}
    \caption{\textit{Top:} The three phase-folded light curves from the bottom row of Figure \ref{fig:Gaia_mep} are shown together on the same axes. We see the difference in amplitudes between each filter for each system type. The $\mathrm{G_{RP}}$ filter shows more variability than the $\mathrm{G_{BP}}$ and $\mathrm{G}$ filters for the reflection effect and HW Vir-type system in (a) and (b) respectively. During eclipses, the flux drops similarly in all three filters. The ellipsoidal system shows consistent variations across all filters over an orbit. \textit{Bottom:} The color curves for each system are folded over the dominant peak in the periodogram when multiplying all three periodograms from the bottom row of Figure \ref{fig:Gaia_mep} together. We see sinusoidal variations in phase with the light curves for the reflection effect and the eclipsing HW Vir-type systems, while the ellipsoidally modulated system remains approximately color-constant throughout its orbit.}
    \label{fig:color_phase_lcs}
\end{figure}

The temperature of a system is a key piece of information when determining what physical processes are driving brightness variations, as luminosity is determined by the Stefan--Boltzmann law $\mathrm{L}= 4\pi\sigma\mathrm{R^2T^4},$ where $\sigma$ is the Stefan--Boltzmann constant, and R and T are the radius and temperature of the star, respectively. The color of a system is determined by the temperature of its photosphere. With near-simultaneous measurements across all three filters at each epoch, Gaia DR3 MEP provides direct and precise color measurements. Calculating  $\mathrm{G_{BP}-G_{RP}}$ for each epoch gives us a color curve showing how the color of a system changes over time. In the bottom panel of Figure \ref{fig:color_phase_lcs}, $\mathrm{G_{BP}-G_{RP}}$ data are folded over the combined period found through multiplying all three filters' normalized periodograms together.

Because light is re-radiated from the dM/BD companion's hot spot in a reflection effect system, during times of peak brightness the light we see is redder compared to when the hot spot is out of view. The phase-folded light curves in Figures \ref{fig:color_phase_lcs}(a) and (b) vary by a few more percent in the red than the blue, causing the star to appear bluer when faint, and redder when bright. For the reflection effect and HW Vir-type systems, the color curve oscillates sinusoidally, approximately in phase with the overall brightness changes shown in the G filter. However, for the ellipsoidal system, there are no discernible color variations; overall the change is about 0.02 magnitudes, and no coherent pattern appears when folded on the system's period.

Gaia DR3 MEP provides two crucial pieces of information: color curves and G magnitude. With distance measurements also provided by Gaia, we can easily calculate $\mathrm{M_G}$ for many objects. These are the necessary pieces of information needed to create a CMD. Gaia's ability to detect small changes in brightness, as discussed in Section \ref{subsec:LCs}, also yields precise colors as seen in Figure \ref{fig:color_phase_lcs}. The photometric precision provided by Gaia DR3 MEP should be sufficient for detecting subtle oscillations of variable stars in the VECMD.

\subsection{Gaia CMD Motions}

Stars are commonly plotted in the Gaia CMD using static points determined by averaging all photometric measurements for a single filter over time. Using the time-resolved $\mathrm{G_{BP}-G_{RP}}$ color and $\mathrm{M_G}$ information provided by Gaia MEP, we can investigate the motions of variable objects through the CMD by constructing the VECMD introduced in Paper I. These variability paths demonstrate how the system changes in color and luminosity through time. The physical processes driving brightness fluctuations uniquely define each system's path through the CMD.

Gaia DR3 MEP is often polluted by outliers, most often in the $\mathrm{G_{BP}}$ and $\mathrm{G_{RP}}$ filters. To combat this, we follow the two-step sigma clipping procedure described in Paper I. First, we clip outliers above and below $4\sigma$ for each individual filter's full Gaia DR3 MEP. After cutting these points, we fit a line to the remaining epochs in the CMD and test how much each point influences its slope, as measured by the ratio of standard deviations $\sigma_{M_G}/\sigma_{(G_{BP}-G_{RP})}$. Starting with the most deviant point, we remove it and recompute the ratio of standard deviations. If the ratio changes by more than 10\%, the epoch is removed from the data set, as the point significantly alters the slope of all points. We then repeat this process iteratively until no points exceed the threshold, ending with a sigma-clipped data set based on the slope direction. Occasionally, this process will clip some eclipse points. However, the method preserves the overall slope of the line, and with it the physics that sets the direction of the slope. 

\begin{figure*}[h!]
\centering
\begin{minipage}[t]{0.325\textwidth}
    \centering
    \includegraphics[width=\linewidth]{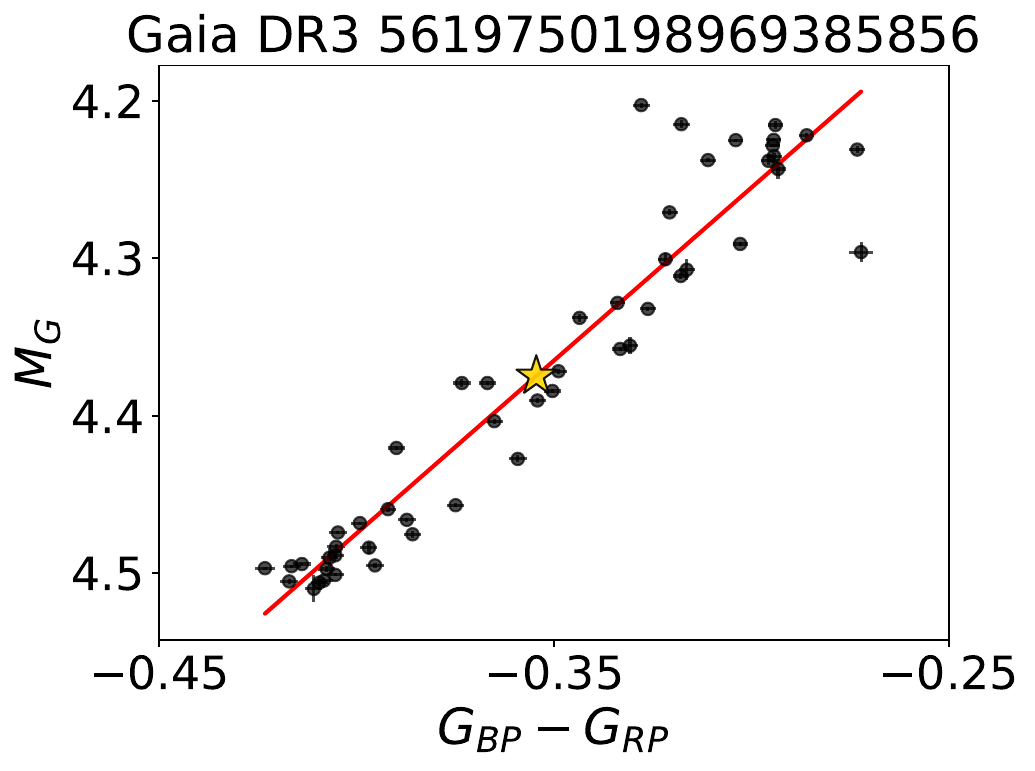}

    \smallskip
    \textbf{(a)}
\end{minipage}\hfill
\begin{minipage}[t]{0.325\textwidth}
    \centering
    \includegraphics[width=\linewidth]{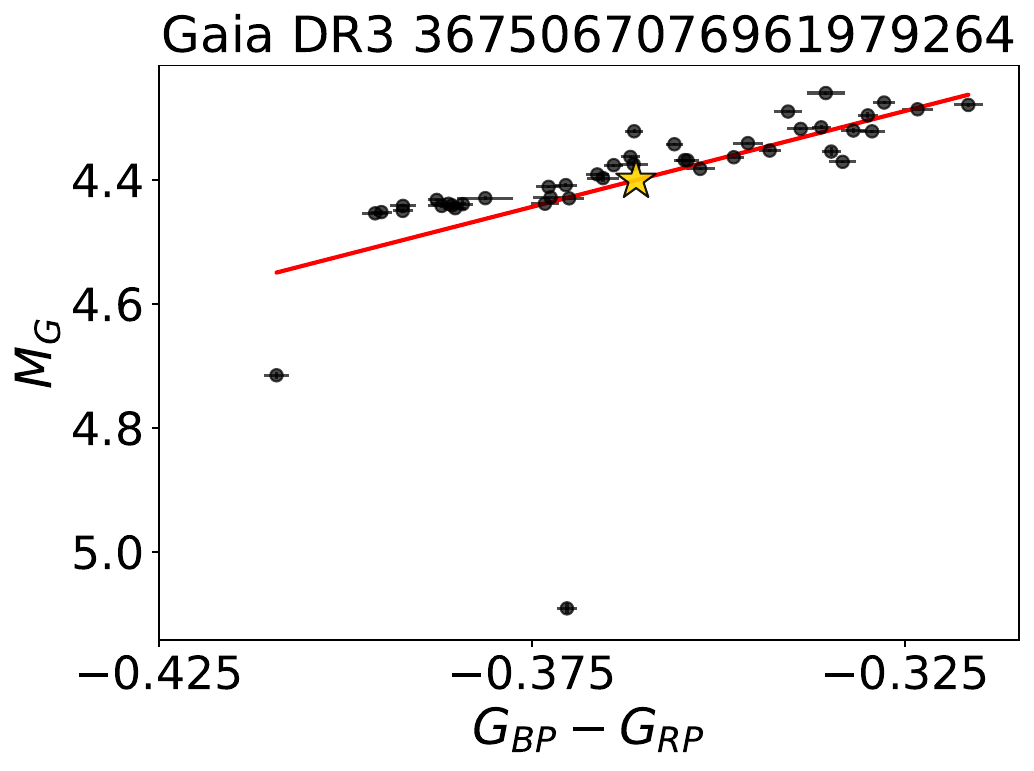}

    \smallskip
    \textbf{(b)}
\end{minipage}\hfill
\begin{minipage}[t]{0.325\textwidth}
    \centering
    \includegraphics[width=\linewidth]{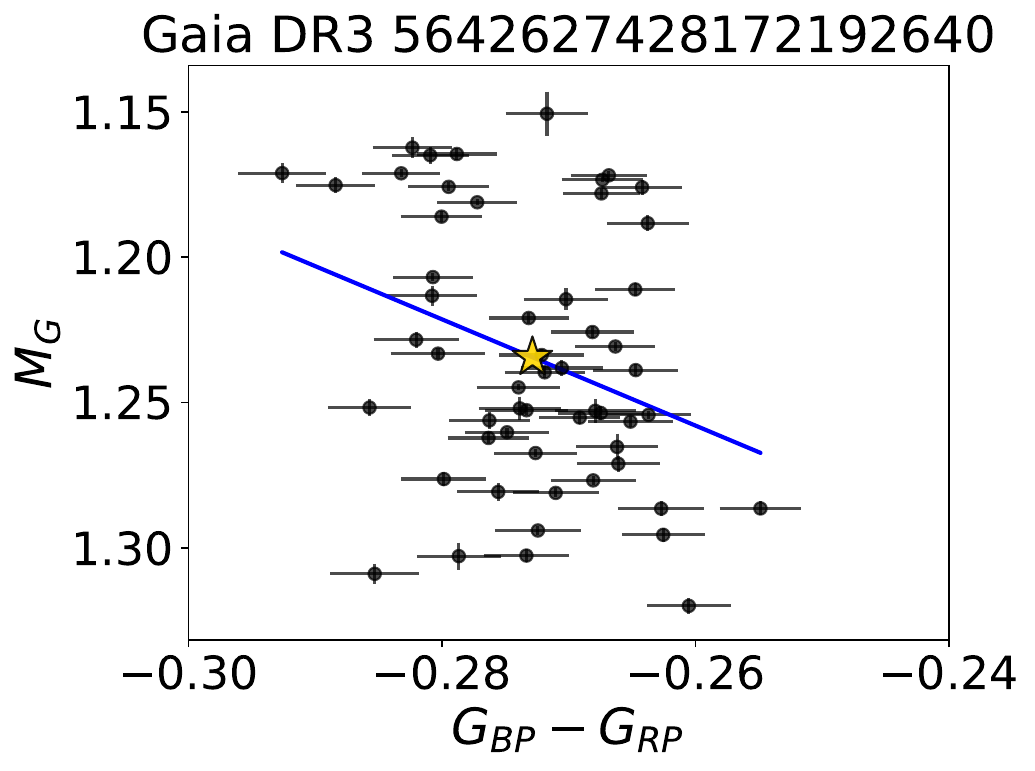}

    \smallskip
    \textbf{(c)}
\end{minipage}

\caption{The variability paths that the reflection effect, HW Vir-type, and ellipsoidal example systems trace out on the Gaia CMD are shown in panels \textbf{(a)}, \textbf{(b)}, and \textbf{(c)}, respectively. With $\mathrm{M_G}$ on the y-axis and $\mathrm{G_{BP}-G_{RP}}$ on the x-axis, individual epochs are plotted on the diagram to see how the position of the object changes over time. The gold star represents the mean of the data, demonstrating how a traditional CMD would represent the system after using our sigma clipping methods. A line is fit to the data to show the slope of the points and colored to indicate whether the system is bluer or redder when brightest. The eclipses in panel \textbf{(b)} drop down to create a hook in the trajectory, while the ellipsoidal system shows much greater variations in overall brightness than in color.}
\label{fig:CMD_plots}
\end{figure*}

Figure \ref{fig:CMD_plots} plots matched $\mathrm{M_G}$ and $\mathrm{G_{BP} - G_{RP}}$ epochs on a Gaia CMD for the reflection effect, HW Vir-type, and ellipsoidal example systems. The amplitudes of the color curves from Figure \ref{fig:color_phase_lcs} are projected onto the x-axis of the CMD, and the G filter light curves from the top panels of Figure \ref{fig:Gaia_mep} are projected onto the y-axis (after a parallax correction). We even see the shape of the light curve manifest in the density of points along the line with clumping near the top right and bottom left in Figure \ref{fig:CMD_plots}(a), just as in the peaks and troughs of Figure \ref{fig:Gaia_mep}(a). In the HW Vir-type system's CMD, the eclipses appear as points dropping below the reflection effect shape. The shape from the ellipsoidal system is close to a vertical dispersion because there are almost exclusively $\mathrm{M_G}$ variations with no color variations. 

Each data set is fit with a line using the \texttt{curve\_fit} function from the \texttt{scipy.optimize} package. By fitting a line, we can show the direction and magnitude of variations in CMD space. A line's color indicates the system's color when it is brightest. For a system where brightness variations are dominated by the reflection effect, the system is brightest when the hot spot is in view, making it appear redder. This method works excellently for a small number of systems in our sample, as there are a limited number of features and shapes we are searching for. However, as noted in Paper I, the simplicity of lines overlooks eclipses and other subtle features such as curvature in the variability paths. For this reason, we always inspect the full CMD trajectories and phase-folded light curves alongside the fitted lines.

\begin{figure}[h!]
    \centering
    \begin{minipage}[c]{0.48\textwidth}
        \centering
        \includegraphics[width=\linewidth]{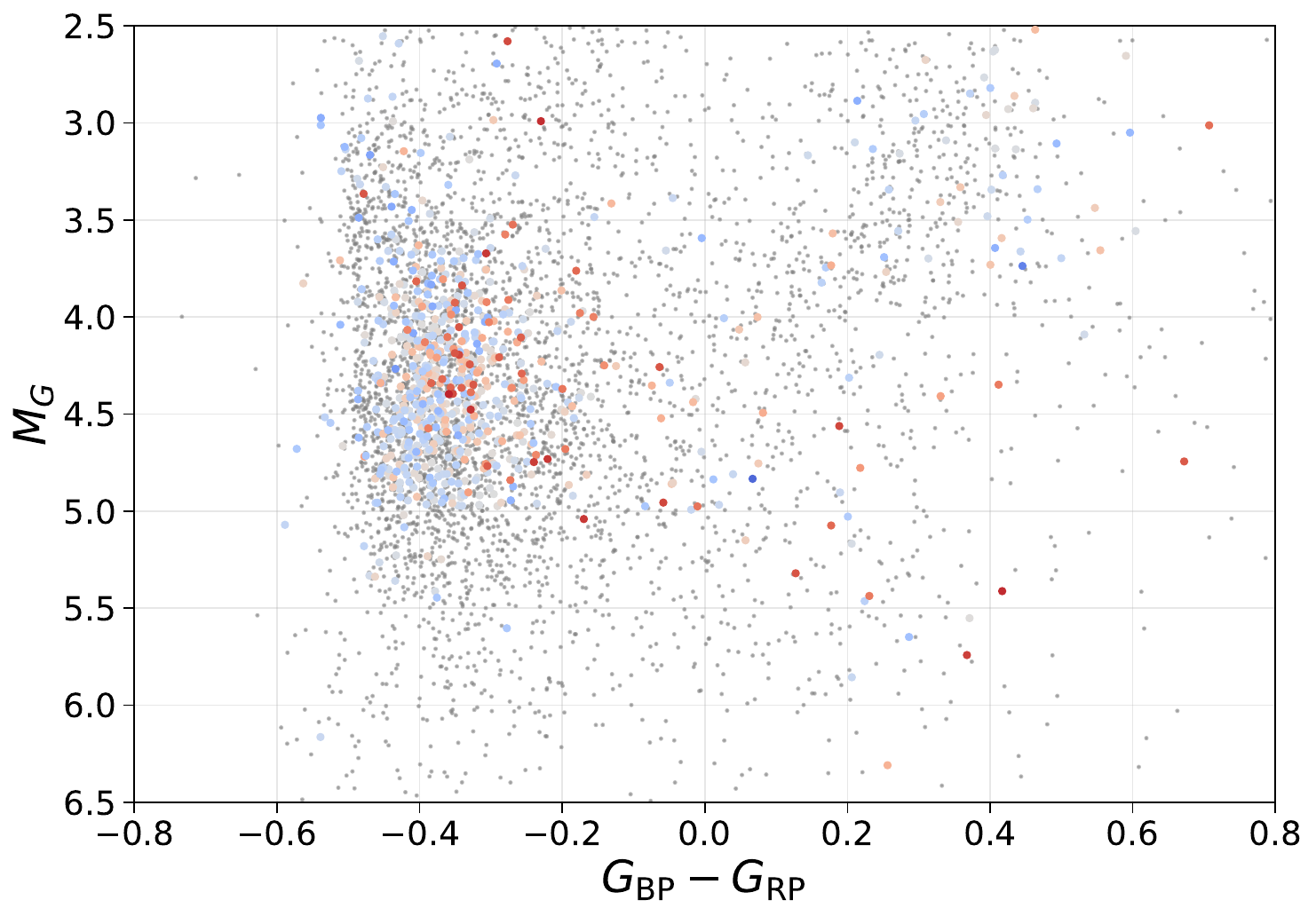}
        \smallskip
        \textbf{(a)}
    \end{minipage}\hfill
    \begin{minipage}[c]{0.48\textwidth}
        \centering
        \includegraphics[width=\linewidth]{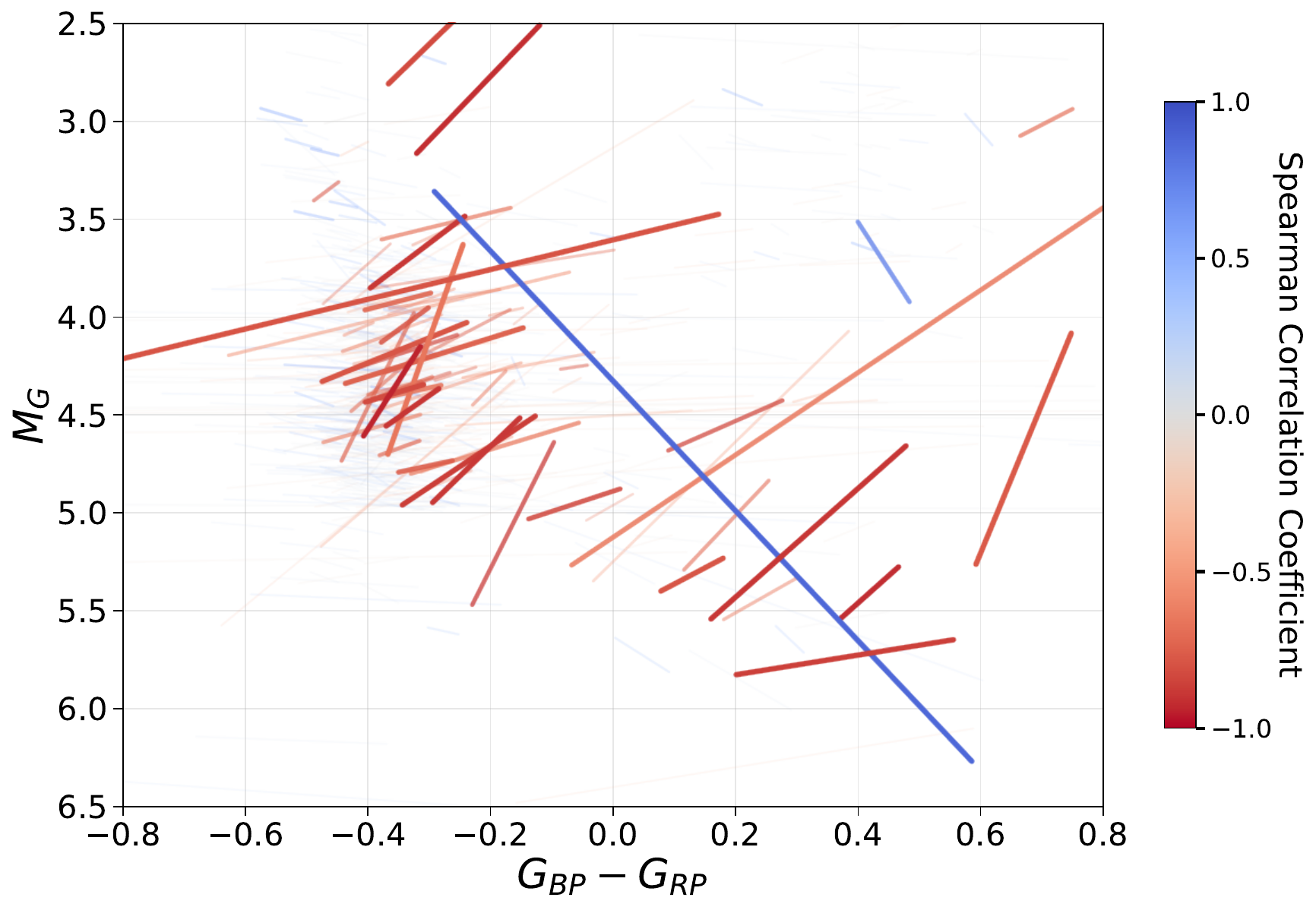}
        \smallskip
        \textbf{(b)}
    \end{minipage}

\caption{The catalog of known sdBs from \citet{Culpan2022} is shown in two different forms. \textbf{(a)} We show a zoomed-in version of a standard CMD. Points with MEP are highlighted with the same color as their corresponding lines in (b) determined by the Spearman correlation coefficient. Note that error bars are not shown on this plot for readability purposes. \textbf{(b)} Points from the known \citet{Culpan2022} catalog with Gaia DR3 MEP available are shown in the VECMD after sigma clipping and fitting lines to individual epochs. The length of the line is the magnitude of the maximum difference in $\mathrm{M_G}$ and $\mathrm{G_{BP}-G_{RP}}$, and the color denotes whether the system is brightest when bluest or brightest when reddest. The color saturation is determined by the strength of the Spearman correlation coefficient for the linear fit and the lines are stacked in order of Spearman correlation p-value with the lowest p-values on top. All of the lines are centered on the mean value for $\mathrm{M_G}$ and $\mathrm{G_{BP}-G_{RP}}$. There is a clump of red lines, all with similar slopes, matching the reflection effect pattern we see in Figure \ref{fig:CMD_plots}. The large blue line is the known cataclysmic variable ``IW And''.}
\label{fig:Known_Culpan}
\end{figure}

We apply this line-fitting procedure to known hot subdwarfs provided by \citet{Culpan2022} to see the dispersion of vectors and the patterns that appear when multiple sources are considered. Of the 6,616 known sdBs in the catalog, 828 have published Gaia DR3 MEP, and these are shown in a standard CMD and in the VECMD in Figure \ref{fig:Known_Culpan}(a) and (b), respectively. To tell how monotonic the trend is in the MEP data, we use a Spearman correlation coefficient, as in Paper I. The Spearman correlation coefficient measures the strength and direction of the rank-order relationship between G magnitude and $\mathrm{G_{BP}-G_{RP}}$ color, capturing whether stars tend to get brighter or fainter as they get redder or bluer. If a linear trend is present in the data, the line will appear darker and overlaid on the fainter lines in the diagram. The red lines with a slope up and to the right, matching the slope we see in both Figure \ref{fig:CMD_plots}(a) and (b), are most likely systems dominated by the reflection effect. These lines all have comparable slopes, demonstrating that stars of the same type that share variability mechanisms will have similar slopes. The almost vertical blue and red lines depicting ellipsoidal systems are fainter, and they show less correlation between epochs. The large blue line cutting through the data is the cataclysmic variable ``IW And''. This outlier clearly stands out from the usual pattern of dark red lines. The red lines whose slopes are greater than others are most likely HW Vir-type systems, as eclipses change the fit of the line to compensate for the lower points. Figure \ref{fig:Known_Culpan} thus provides a template for our search. Compact sdB binaries of a given type share trajectory shapes, which we exploit in Section \ref{sec:target} to search for new systems that are missing from the \citet{Culpan2022} catalogs.

\section{Target Selection}\label{sec:target}

Most sdB studies currently utilize the hot subdwarf catalogs created by \citet{Culpan2022}. They provide a list of 61,585 candidate hot subdwarfs, of which 2,114 objects have released Gaia DR3 MEP \citep{Princy2025}. To create this list, they started with a CMD cut around the EHB and applied astrometric and quality criteria. Most notably, in their CMD cut, they considered only systems with parallax uncertainties smaller than 20\%. While strict cuts are essential to providing a scrupulous search for sdB candidates based on CMD position alone, they exclude any system whose parallax is too imprecise to place it confidently on the EHB. As shown in Paper I and Section \ref{sec:MEP}, however, the CMD trajectory of a variable sdB is set by the physics of its variability, not by its parallax. Therefore, we deliberately searched {\em only} the systems that fall within the CMD region used by \citet{Culpan2022} but did not fulfill their astrometric and quality requirements, using light curve morphologies and variability paths in the VECMD to identify new sdBs. 

\begin{figure}[h!]
    \centering
    \begin{minipage}[c]{0.48\textwidth}
        \centering
        \includegraphics[width=\linewidth]{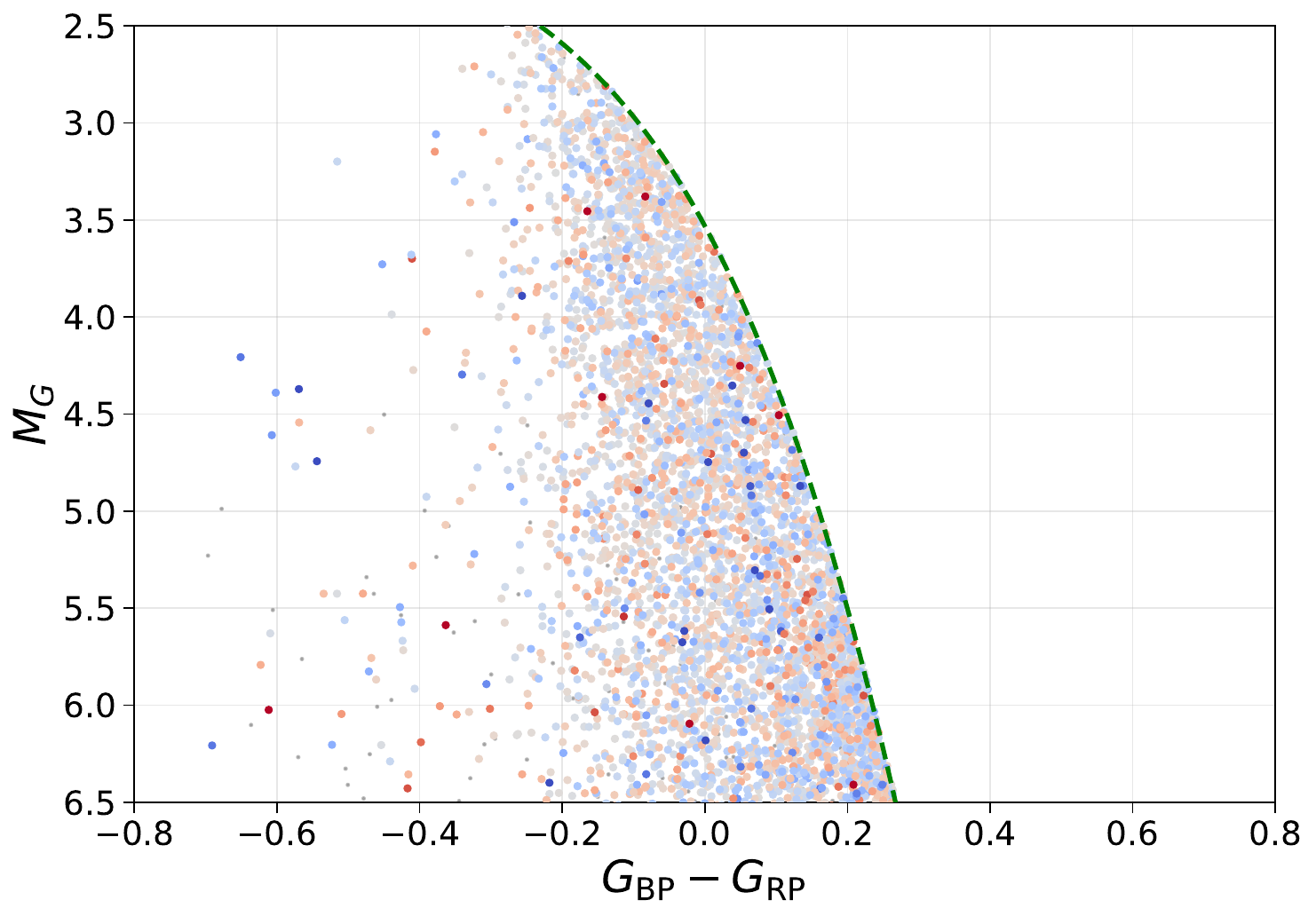}
        \smallskip
        \textbf{(a)}
    \end{minipage}\hfill
    \begin{minipage}[c]{0.48\textwidth}
        \centering
        \includegraphics[width=\linewidth]{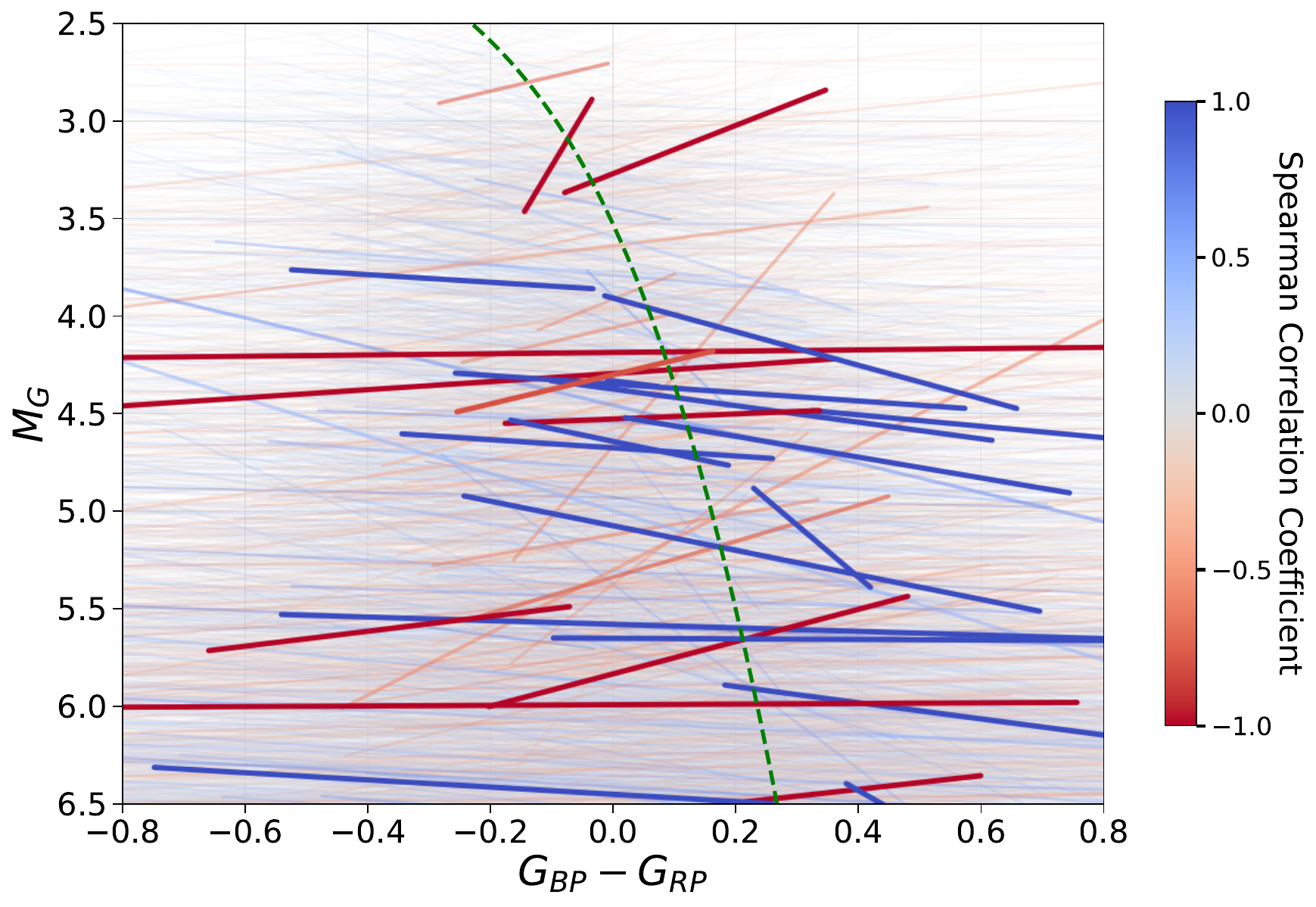}
        \smallskip
        \textbf{(b)}
    \end{minipage}

\caption{A set of 5,782 unknown systems is shown in a similar form as in Figure \ref{fig:Known_Culpan}. \textbf{(a)} We see a polynomial cutoff on the right denoted by the dashed green line and a collection of points to the left of this. Many of these points have highly uncertain parallaxes, making their exact positions in the CMD poorly defined. Note that error bars are not shown on this plot for readability purposes. \textbf{(b)} We see 5,604 variability paths of the 5,782 total points plotted as lines on the VECMD after they have gone through our sigma clipping process. Out of our original data set, 178 systems were not able to be plotted as vectors in the VECMD due to extremely poor data, excessive sigma clipping removing too many points, or fewer than two $\mathrm{G_{BP}-G_{RP}}$ points. The points that are not able to be represented with vectors remain as small gray points in (a). Many of these sources are extremely faint or lie in crowded fields and have unreliable MEP, making many of the strong lines untrustworthy. We searched by eye through all 5,782 systems to identify variability paths and corresponding light-curve morphologies that match known systems.}
\label{fig:Anti-Culpan}
\end{figure}

We started by taking every star deemed to fall within the ``strict hot subluminous star selection criterion'' that \citet{Culpan2022} used to separate the hot subluminous stars from the main sequence. We chose the stricter area to ensure these stars were some sort of hot subluminous star rather than main sequence stars that would pollute our sample. This was one of the main criterion \citet{Culpan2022} used when determining candidates solely based on their position in the CMD. Also, we only utilized systems with positive parallax so that each system could be plotted on a CMD. We placed no constraint on the precision of the parallax. These cuts yielded 797,502 systems, which were then cross-referenced with the known and candidate sdB lists to remove any systems that may have been included in either catalog. This left us with 769,609 systems, none of which would appear in a search built on the standard hot subdwarf catalogs. Next, we cross-referenced with Gaia DR3 MEP files, yielding a final total of 5,782 uncataloged objects with Gaia MEP shown in Figure \ref{fig:Anti-Culpan}(a). We show the variability paths for these objects in Figure \ref{fig:Anti-Culpan}(b).

With this method, the slope of the line and high Spearman correlation coefficient help determine whether the system exhibits reflection effect, HW Vir-type, or ellipsoidal variability. Unlike the criteria imposed in \citet{Culpan2022} to identify sdBs based on their position in the CMD, knowing the exact vertical position of the sdB in the VECMD is not needed, since an imprecise parallax only shifts its trajectory up or down without changing its shape. We can identify candidate sdBs and classify those without well-defined astrometric parameters using the shape of the variability paths on the VECMD and the phase-folded Gaia DR3 MEP light and color curves. There are some lines that stand out in Figure \ref{fig:Anti-Culpan} as very high correlation systems, but these sources only have about $2\ \mathrm{G_{BP}-G_{RP}}$ points matched with G points. When a line is fit to these epochs, the Spearman correlation coefficient is very close to $\pm1$, giving a strong line when there isn't enough data to make a classification. The dashed green line in both plots denotes the polynomial cutoff from \citet{Culpan2022} used to gather our list of targets. Some lines sit to the right of the polynomial cutoff because our sigma clipping algorithm removes points, shifting their average position to the left or right of the traditional CMD point.

We examined each phase-folded light curve, color curve, CMD motion plot, and full Gaia DR3 MEP photometry for all 5,782 systems by eye. These figures included the full Gaia DR3 MEP photometry in all three filters (G, $\mathrm{G_{BP}}$, and $\mathrm{G_{RP}}$) with a computed color curve ($\mathrm{G_{BP}} - \mathrm{G_{RP}}$), the phase-folded MEP data as seen in Figure \ref{fig:Gaia_mep} with the phase-folded color curve as in Figure \ref{fig:color_phase_lcs} both folded over the strongest peak and twice the strongest peak, and the CMD motion plot as seen in Figure \ref{fig:CMD_plots}. Candidates required all plots to be consistent with known sdB variables. If a system demonstrated a reflection effect light curve shape, we also phase-folded the system with a standard $4\sigma$ clipping algorithm applied only above the data to check for missed eclipses.



\section{New Variable Hot Subdwarfs}\label{sec:results}

Using the CMD trajectories from Gaia DR3, we found 27 new variable sdB candidates, including 12 candidate reflection effect binaries and 15 HW Vir-type systems, as listed in Table \ref{tab:results}. We also recovered one previously known sdB, discussed below. These new HW Vir-type candidates represent a substantial addition to the known population of eclipsing sdB binaries. Importantly, most are new sdB candidates altogether. Four of the 27 systems (marked with $\dagger$ in Table \ref{tab:results}) appeared in the Gaia DR2 catalog of hot subluminous star candidates by \citet{Geier2019} but were not carried into the \citet{Culpan2022} catalogs. The remaining 23 systems had never been identified as hot subdwarfs.  We also confirm the findings of \citet{Koen2023}, who identified Gaia DR3 4488370193035250432 in ATLAS data as a reflection effect sdB.

Including our new sdB variables, many systems were flagged as variable in at least one of the three filters. Many of these systems could not be phase-folded, had a poor signal-to-noise ratio, or were not periodic. While these systems are important and many remain unclassified, here we report only new sdB variables that we can preliminarily classify using light curve and variability path shapes. In Section \ref{sec:conclusion}, we will discuss future plans utilizing all systems. We stress that the classifications presented here are preliminary and based solely on Gaia photometry. Spectroscopic follow-up is required to confirm that these systems are bona fide hot subdwarfs, and to measure their fundamental properties. Additionally, it is important to note that we did not find any clear ellipsoidal systems in this search due to the sensitivity of the measurements and the precision needed to detect the difference in the peaks and troughs. 

\begin{deluxetable}{ccccccc}[h!]
\tablecaption{New Candidate sdB Binaries}
\label{tab:results}
\tablehead{
\colhead{Gaia DR3} &
\colhead{RA (J2016)} &
\colhead{DEC (J2016)} &
\colhead{Subtype} &
\colhead{Avg G} &
\colhead{Avg $\mathrm{G_{BP}-G_{RP}}$} &
\colhead{Period (d)}
}
\startdata
4488370193035250432\textsuperscript{*} & 17:48:37.36 & +09:09:48.15 & Reflection & 16.62 & $-$0.18  & 0.09646 \\
1994949881653072128 & 23:50:16.20 & +56:24:36.78 & Reflection & 18.47 & 0.24   & 0.11147 \\
5824146288167395200 & 15:33:30.99 & $-$66:12:52.24 & Reflection & 16.82 & $-$0.19  & 0.11710 \\
5360524321087229952 & 10:46:01.40 & $-$51:45:32.83 & Reflection & 17.02 & 0.16   & 0.12142 \\
4295932224878864640 & 19:30:44.22 & +07:53:13.11 & Reflection & 16.37 & -0.01  & 0.13844 \\
5936327771600351104\textsuperscript{\dag} & 17:01:06.90 & $-$52:25:12.24 & Reflection & 17.30 & 0.01   & 0.15338 \\
4130984211988790144 & 16:35:54.08 & $-$20:09:31.65 & Reflection & 16.63 & 0.09   & 0.15903 \\
2071824122597458048 & 19:55:06.70 & +36:51:45.92 & Reflection & 18.83 & 0.10   & 0.17815 \\
2071866144570594304 & 19:56:27.66 & +37:27:02.16 & Reflection & 17.27 & 0.04   & 0.19627 \\
4250303415695312128\textsuperscript{\dag} & 20:08:59.23 & +06:24:57.20 & Reflection & 16.43 & $-$0.11  & 0.21845 \\
4303099116530878208\textsuperscript{\dag} & 19:42:57.44 & +11:27:44.35 & Reflection & 17.07 & $-$0.07  & 0.23030 \\
5235621350686875776 & 11:58:47.24 & $-$67:58:31.28 & Reflection & 16.47 & $-$0.06  & 0.25842 \\
4092234394269347840 & 18:42:55.41 & $-$19:51:18.14 & Reflection & 15.94 & 0.08   & 0.39170 \\
\tableline
547873656141273728  & 03:15:23.62 & +74:49:54.07 & HW Vir     & 17.01 & $-$0.02  & 0.09157 \\
4504497623459700992 & 18:48:47.06 & +11:57:20.21 & HW Vir     & 16.69 & 0.06   & 0.10683 \\
1818797810959434496 & 20:34:30.53 & +22:42:25.11 & HW Vir     & 17.59 & 0.02   & 0.12259 \\
5317568154136012928 & 08:51:11.09 & $-$53:38:52.39 & HW Vir     & 18.31 & 0.11   & 0.12799 \\
5339944353944771712 & 11:14:17.92 & $-$57:25:04.86 & HW Vir     & 16.64 & 0.09   & 0.14115 \\
5253478141071678464 & 10:39:01.93 & $-$62:54:24.89 & HW Vir     & 18.00 & 0.06   & 0.16538 \\
4317534707775792896 & 19:49:06.81 & +14:53:03.76 & HW Vir     & 16.92 & 0.02   & 0.20048 \\
5852569385332097920 & 14:22:07.32 & $-$65:35:20.29 & HW Vir     & 16.92 & 0.11   & 0.21037 \\
5869649194601275776 & 13:20:18.45 & $-$60:19:51.82 & HW Vir     & 17.22 & 0.12   & 0.22174 \\
4316985776556532736 & 19:38:48.73 & +12:49:52.65 & HW Vir     & 18.07 & 0.14   & 0.23750 \\
5954658863810768640\textsuperscript{\dag} & 17:52:19.03 & $-$44:59:32.39 & HW Vir & 16.58 & $-$0.17  & 0.25091 \\
4687195838491561728 & 01:06:32.86 & $-$73:15:31.68 & HW Vir     & 19.34 & $-$0.09  & 0.31494 \\
5932001536935347456 & 16:27:27.71 & $-$54:15:24.44 & HW Vir     & 17.63 & 0.02   & 0.32686 \\
6060578877657380736 & 12:57:31.21 & $-$57:20:11.29 & HW Vir     & 19.10 & 0.22   & 0.33711 \\
5921949393628901248 & 17:38:59.40 & $-$52:10:45.59 & HW Vir     & 16.87 & 0.00   & 0.35854 \\
\enddata
\tablecomments{
\textsuperscript{*}Previously known reflection effect sdB, independently recovered here \citet{Koen2023};
\textsuperscript{\dag}Hot subluminous candidate in Gaia DR2 \citet{Geier2019}
}
\end{deluxetable}

\begin{figure}[p!]
    \centering
    \includegraphics[width =0.87\linewidth]{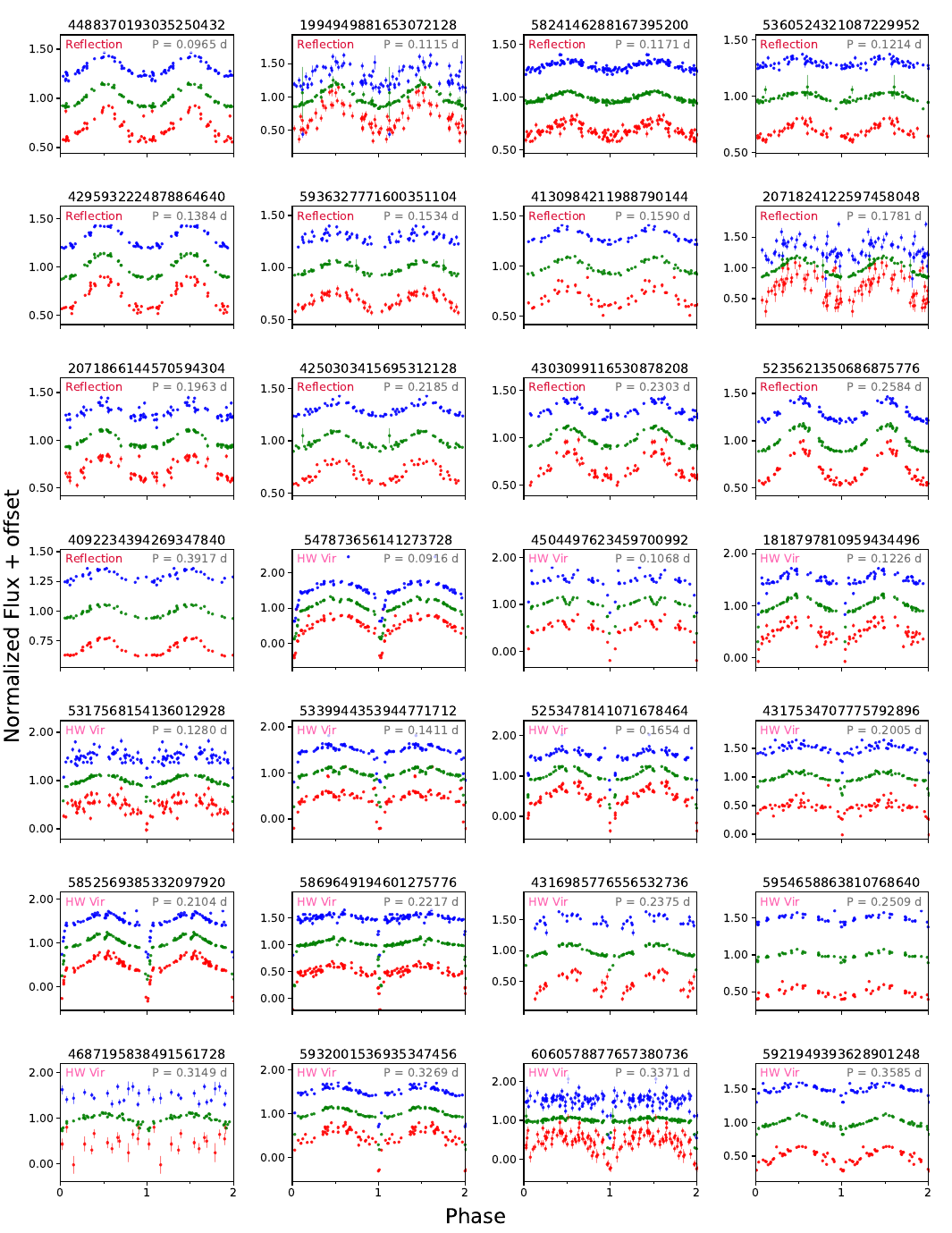}
    \caption{All Gaia MEP light curves for systems presented in Table \ref{tab:results} are presented in the same form as the top row of Figure \ref{fig:color_phase_lcs}. Systems are folded over the combined periodogram except for Gaia DR3 4504497623459700992, Gaia DR3 4687195838491561728, Gaia DR3 5936327771600351104, Gaia DR3 4316985776556532736, and Gaia DR3 5869649194601275776, which are phase-folded over only the G filter as the combined periodogram was incoherent due to poor $\mathrm{G_{BP}}$ and $\mathrm{G_{RP}}$ measurements. We see a wide range of strengths for the reflection effect as well as deep and shallow eclipses for HW Vir-type systems.}
    \label{fig:ALL_LCs}
\end{figure}

Figure \ref{fig:ALL_LCs} shows all of the phase-folded light curves together. Many HW Vir-type systems show primary eclipses that drop below 50\% of the relative flux, and many reflection effect systems demonstrate a semi-amplitude $> 10\%$. The shortest-period HW Vir-type system discovered was Gaia DR3 547873656141273728 with a period of $0.09157$ days ($2.198$ hours), and the longest-period HW Vir-type was Gaia DR3 5921949393628901248 with a period of $0.35854$ days ($8.605$ hours). Interestingly, we find four HW Vir-type systems with a detected orbital period of over seven hours (Gaia DR3 4687195838491561728, Gaia DR3 5932001536935347456, Gaia DR3 6060578877657380736, and Gaia DR3 5921949393628901248), with two of those systems having  eclipse depths of over $60\%$ (Gaia DR3 5932001536935347456 and Gaia DR3 6060578877657380736). The brightest HW Vir-type system has a magnitude of $16.58$ in the G filter, while the dimmest has a magnitude of $19.34$. Additionally, 21 of the 27 systems lie within 10$^\circ$ of the Galactic plane, demonstrating that this methodology can reach crowded and reddened regions of the Galaxy, where astronomers have had difficulty finding sdB systems \citep{Culpan2022}. Among the new candidate reflection effect systems, the shortest-period system is Gaia DR3 1994949881653072128 with a period of 0.11147 days (2.675 hours), and the longest is Gaia DR3 4092234394269347840 with a period of 0.39170 days (9.401 hours). The largest peak-to-peak reflection amplitude is $\sim$30\%. The magnitudes of these systems range from $15.94$ to $18.83$ in the G filter. 

\FloatBarrier

\section{Summary and Conclusion}\label{sec:conclusion}

We demonstrate a methodology for finding new sdBs using Gaia DR3 MEP, based on the shape and slope of $\mathrm{M_G}$ and $\mathrm{G_{BP}-G_{RP}}$ epochs in the VECMD introduced in Paper I. Because the shape of a CMD trajectory is set by the physics driving a system's variability rather than by its parallax, this approach does not require the astrometric and crowding cuts through which standard hot subdwarf catalogs are built. We applied it to 5,782 systems that lie in the hot subdwarf region of the Gaia CMD but were left out of the \citet{Culpan2022} catalogs, and identified 27 new candidate compact sdB binaries, including 15 HW Vir-type systems and 12 reflection effect systems. Since 23 of these systems were not previously known to be hot subdwarfs, our search has produced new hot subdwarf candidates otherwise overlooked by prior catalog cuts. In future work, this could allow us to extend the search for sdBs further into the main sequence, as they will stand out as outliers in the overall stagnant main sequence. Additionally, this method can be applied to other variable stars, such as those along the WD cooling sequence and pulsators. 

The classifications presented here are based solely on Gaia photometry. Higher signal-to-noise light curves and spectroscopic follow-up are required to confirm that these systems are hot subdwarfs, determine their atmospheric parameters, and measure their radial velocity curves. Nonetheless, the light curve morphologies and CMD trajectories of these systems closely match those of known reflection effect and HW Vir-type sdB binaries, and we consider them strong candidates.

We are currently missing targets that lack clean phase-folded light curves, as we cannot fully classify them. Some systems exhibited sinusoidal oscillations but were not sufficiently distinct to be classified as candidate reflection effect or ellipsoidal systems. Additionally, many targets lacked enough data points to fully see the slope on the VECMD and make a classification. The new data from Gaia DR4, scheduled for release on 2026 December 2, will provide MEP for many more systems and additional epochs for systems already released in DR3, which should allow us to identify ellipsoidal sdBs and classify previously indistinct sinusoidal curves using our methodology.

Because our classification relies on shapes, machine learning should be highly effective for object classification. Adding these new parameters and shapes to machine learning algorithms will greatly enhance their performance and enable the rapid classification of large numbers of objects. We have already begun this work and will analyze the entire Gaia DR3 MEP catalog and develop a pipeline for Gaia DR4 MEP.

\begin{acknowledgments}
This project is supported by the College of Arts and Sciences' Seed Funding Grant program at the University of North Carolina at Chapel Hill and the Wells Family. Additionally, B.K.  acknowledges support from the Summer Undergraduate Research Fellowship (SURF) program.
This work has made use of data from the European Space Agency (ESA) mission {\it Gaia} (\url{https://www.cosmos.esa.int/gaia}), processed by the {\it Gaia} Data Processing and Analysis Consortium (DPAC, \url{https://www.cosmos.esa.int/web/gaia/dpac/consortium}). Funding for the DPAC has been provided by national institutions, in particular the institutions participating in the {\it Gaia} Multilateral Agreement. This paper includes data collected with the TESS mission, obtained from the MAST data archive at the Space Telescope Science Institute (STScI). Funding for US Institutions for the TESS mission is provided by the NASA Explorer Program. STScI is operated by the Association of Universities for Research in Astronomy, Inc., under NASA contract NAS5–26555.

\end{acknowledgments}

%

\pagebreak
\facilities{Gaia, TESS}


\software{Astropy \citep{astropy:2013, astropy:2018, astropy:2022}, Matplotlib \citep{Hunter2007}, Pandas \citep{mckinney-proc-scipy-2010}, NumPy \citep{Harris2020}, SciPy \citep{2020SciPy-NMeth}}



\bibliography{Main}{}
\bibliographystyle{aasjournal}



\end{document}